\documentclass[reprint,
superscriptaddress,
 amsmath,amssymb,
prb,
]{revtex4-2}

\usepackage{graphicx}
\usepackage{dcolumn}
\usepackage{bm}
\usepackage{comment}
\usepackage{xcolor}

\newcommand*{\hatH}{\hat{\mathcal{H}}}

\begin{document}

\preprint{APS/123-QED}

\title{NMR Investigation on Honeycomb Iridate Ag$_3$LiIr$_2$O$_6$ }

\author{Jiaming Wang}
\affiliation{%
 Department of Physics and Astronomy, McMaster University, Hamilton, Ontario, L8S 4M1, Canada
}%
\author{Weishi Yuan}
\affiliation{%
 Department of Physics and Astronomy, McMaster University, Hamilton, Ontario, L8S 4M1, Canada
}%
\author{Takashi Imai}

\affiliation{%
 Department of Physics and Astronomy, McMaster University, Hamilton, Ontario, L8S 4M1, Canada
}%
\author{Philip M.Singer}
\affiliation{
 Department of Chemical and Biomolecular Engineering, Rice University, Houston, Texas, 77005, USA
}%
\author{Faranak Bahrami}
\affiliation{
 Department of Physics, Boston College, Chestnut Hill, Massachusetts 02467, USA
}%
\author{Fazel Tafti}
\affiliation{
 Department of Physics, Boston College, Chestnut Hill, Massachusetts 02467, USA
}%

\date{\today}

\begin{abstract}

Ag$_3$LiIr$_2$O$_6$ is a Kitaev spin liquid candidate material synthesised from $\alpha$-Li$_2$IrO$_3$ through a topotactic reaction. We investigate the structural and magnetic properties of two samples of Ag$_3$LiIr$_2$O$_6$ based on $^7$Li nuclear magnetic resonance line shape, Knight shift and spin-lattice relaxation rate $1/T_1$. The first sample A shows signatures of magnetically ordered spins, and exhibits one sharp $^7$Li peak with FWHM increasing significantly below 14~K. $1/T_1^{stretch}$ of this sample displays a broad local maximum at 40~K, followed by a very sharp peak at $T_N = 9\pm1$~K due to critical slowing down of Ir spin fluctuations, a typical signature of magnetic long range order.  In order to shed light on the position-by-position variation of $1/T_1$ throughout the sample, we use numerical Inverse Laplace Transform $T_1$ (ILTT$_1$) analysis based on Tikhonov regularization to deduce the density distribution function $P(1/T_1)$.  We demonstrate that $\sim 60\%$ of Ir spins are statically ordered at the NMR measurement timescale but the rest of the sample volume remains paramagnetic even at 4.2~K, presumably because of structural disorder induced primarily by stacking faults. In order to further investigate the influence of structural disorder, we compare these NMR results with those of a second sample B, which has been shown by transmission electron microscope to have domains with unwanted Ag inclusion at Li and Ir sites within the Ir honeycomb planes. The sample B displays an additional NMR peak with relative intensity of $\sim 17\%$. The small Knight shift and $1/T_1$ of these defect-induced $^7$Li sites and the enhancement of bulk susceptibility at low temperatures suggest that these defects generate domains of only weakly magnetic Ir spins accompanied by free spins, leading to a lack of clear signatures of long-range order. The apparent lack of long-range order could be easily misinterpreted as evidence for the realization of a spin liquid ground state in highly disordered Kitaev lattice.

\end{abstract}

\maketitle


\section{Introduction}

In contrast to conventional magnetism, quantum spin liquids (QSLs) are an exotic state of matter which avoid magnetic order even at absolute zero. They instead have a highly entangled ground state induced by frustrated magnetic interactions \cite{Balents2010}. In 2006, Alexander Kitaev demonstrated analytically that a honeycomb lattice of effective spin 1/2 moments interacting via bond-dependent Ising interactions $\hatH_{K}$ has a spin liquid ground state\cite{Kitaev2006}. This is a promising class of QSLs as it can manifest in certain Mott insulators with strong spin-orbit coupling \cite{Jackeli_2009}.

Since then, many materials have been shown to exhibit strong Kitaev interactions, providing opportunities for achieving a Kitaev QSL ground state via chemical tuning. Thus, searching for materials with Kitaev honeycomb planes has become the subject of extensive research \cite{Takagi2019}. Various Kitaev QSL candidates with magnetic transition metal ions have been identified, including Na$_2$IrO$_3$\cite{Singh2010,Simutis2018,Sarkar2020}, $\alpha$-Li$_2$IrO$_3$ \cite{Singh2012}, Li$_2$RhO$_3$ \cite{Khuntia2018}, $\alpha$-RuCl$_3$ \cite{Plumb2016,Sandilands2015,Baek2017,Zheng2017,Jansa2018,Nagai2020}, H$_3$LiIr$_2$O$_6$ \cite{Kitagawa2018}, Cu$_2$IrO$_3$ \cite{Abramchuk2017,takahashi2019,Kenny2019}, and D$_3$LiIr$_2$O$_6$ \cite{Geirhos2020}.

However, additional interactions such as the isotropic Heisenberg interaction $\hatH_{H}$ and symmetric off-diagonal exchange term $\hatH_{\Gamma}$ compete with the frustrated Ising interaction $\hatH_{K}$ and mask the intrinsic Kitaev behaviour \cite{Chaloupka_2010,rau2016}. For example, materials such as Na$_2$IrO$_3$ \cite{ChaloupkaPRL2013,Liu2018} and $\alpha$-RuCl$_3$ \cite{Cao2016,Sears2015} undergo long-range antiferromagnetic (AF) order at $T_{N} \simeq 17$~K or below.

\begin{figure}
\centering
\includegraphics[width=\linewidth]{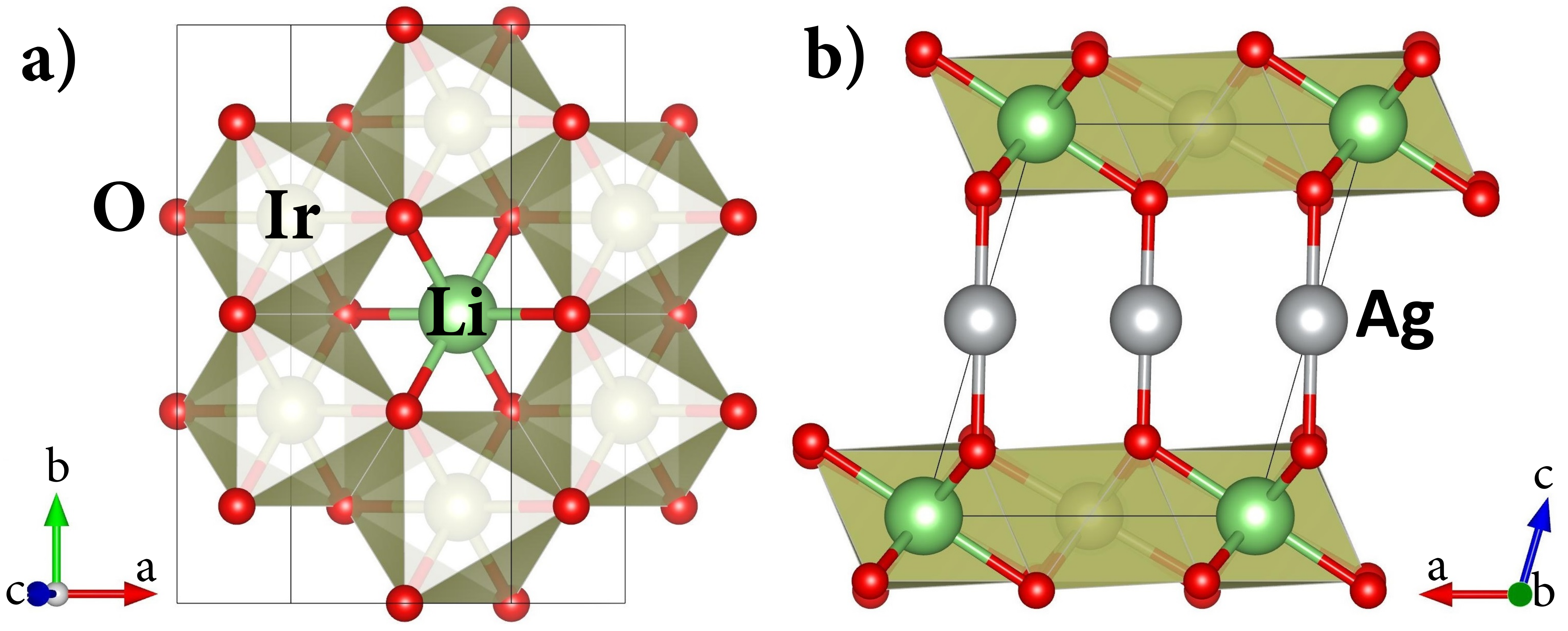}
\caption{a) Ir$^{4+}$ Honeycomb planes in Ag$_3$LiIr$_2$O$_6$ crystal structure \cite{Bahrami2019,Bette2019}. Each Ir$^{4+}$ ion is surrounded by 6 O$^{2-}$ ions (red), forming an octahedron (highlighted in tan). Li$^{+}$ (green) ions are situated at the center of the Ir hexagons. b) Inter-layer Ag$^+$ ions (grey) in-between honeycomb planes. The grey lines indicate the unit cell of the lattice.}
\label{crystal}
\end{figure}

One approach to tune the relative strength of $\hatH_{K}$, $\hatH_{H}$, and $\hatH_{\Gamma}$ is to modify the intra-layer orbital overlaps and bond angles by changing the inter-layer bonding. This was done in $\alpha$-Li$_2$IrO$_3$ by replacing the inter-layer Li$^+$ ions with Ag$^+$ ions to produce Ag$_3$LiIr$_2$O$_6$ \cite{Bahrami2019}, the lattice of which is shown in Fig.\ref{crystal}. Heat capacity and magnetic susceptibility measurements conducted for early samples with significant disorder due both to stacking faults and unwanted Ag-inclusion within the Li and Ir sites in the Kitaev layers have shown no evidence for AF long-range order. However, more recent muon spin rotation ($\mu$SR) measurements conducted on a cleaner sample without inclusion of Ag at the in-plane sites exhibits a Bessel function type oscillation below $T_N \sim 8$~K, establishing incommensurate AF ordering for the majority of the sample volume \cite{Bahrami2021}.

In this report, we use $^7$Li NMR to probe the local structural and magnetic properties of the honeycomb layers in Ag$_3$LiIr$_2$O$_6$, and the potential roles played by structural disorder. We also investigate the spatial distribution of Ir spin dynamics by probing the density distribution function $P(1/T_{1})$ of the $^7$Li NMR spin-lattice relaxation rate $1/T_{1}$ based on inverse Laplace transform (ILT) $T_1$ analysis technique \cite{Singer2020}.  We will demonstrate that $\sim60$~\% of the sample volume undergoes antiferromagnetic long range order below $T_{N} = 9 \pm 1$~K, but the rest of the volume remains paramagnetic even in the cleanest sample without the unwanted Ag-inclusion.  Comparison with a disordered sample indicates that the NMR as well as thermodynamic signatures of the long range order could be easily washed away in more disordered samples.

\section{Experimental Methods}

\begin{figure}[b!]
\centering
\includegraphics[width=0.95\linewidth]{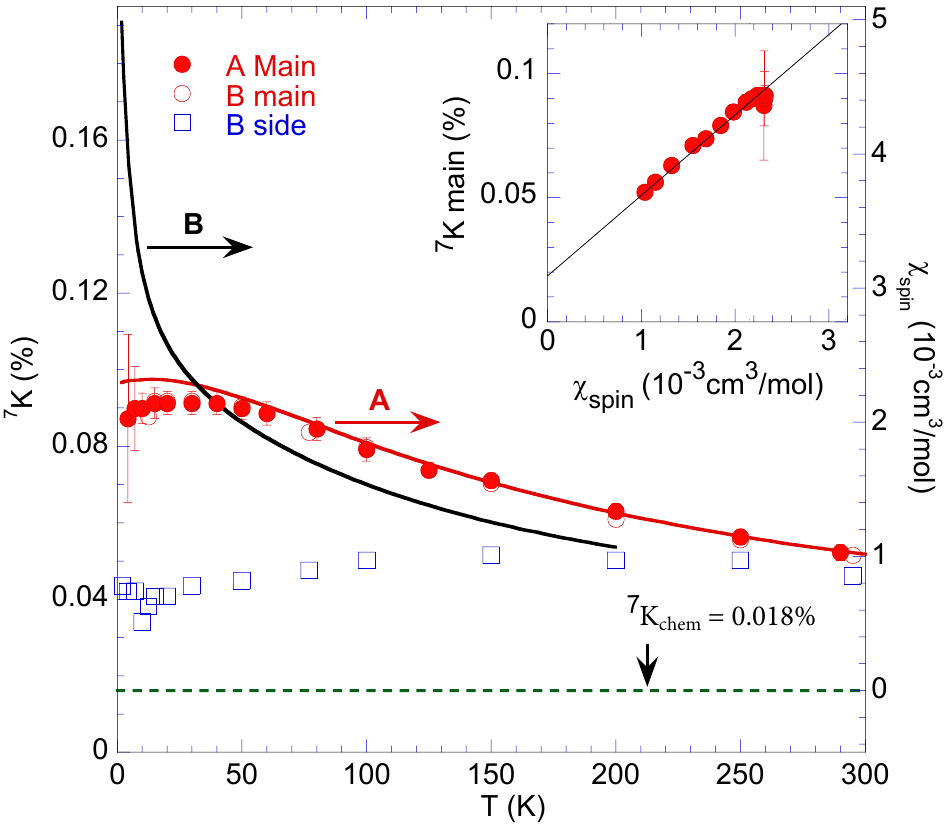}
\caption{Knight shift of main peak for sample A (filled red bullets) along with main (open red circles) and side (open blue squares) peaks for sample B. The bulk spin susceptibility is plotted as the red line and black line for sample A and B, respectively.  The dashed green horizontal line indicates the chemical shift $^7$K$_{chem} \simeq 0.018\%$ of sample A.  The vertical axis of $\chi_{\text{spin}}$ is scaled in proportion to the Knight shift of sample A, and its origin is vertically shifted to account for the chemical shift of sample A. The inset shows Knight shift plotted against spin susceptibility $\chi_{\text{spin}}$ by choosing temperature as their implicit parameter.}
\label{knightshift}
\end{figure}

Powder samples of Ag$_3$LiIr$_2$O$_6$ were synthesised for NMR measurements via a topotactic reaction as described in Ref. \cite{Bahrami2019,Bahrami2021}. We present NMR investigation of two distinct samples, which we refer to as samples A and B. Sample B was synthesized in the early days~\cite{Bahrami2019}, and has inclusions of unwanted Ag ions occupying the Li and Ir sites within the honeycomb plane, as evidenced by clusters observed in transmission electron microscope (TEM) measurements \cite{Bahrami2021}.  This is the sample that appeared to be a spin liquid material~\cite{Bahrami2019}, because the thermodynamic~\cite{Bahrami2019}, $\mu$SR~\cite{Bahrami2019}, and NMR measurements (this work) do not reveal evidence for magnetic long range order.  Furthermore, sample B exhibits an extra split off peak in $^{7}$Li NMR lineshape (as shown below), indicating that unwanted Ag clusters lead to a different type of Li sites.  On the other hand, sample A is a newer sample obtained after the sample synthesis conditions have been refined, and exhibit the signature of magnetic long range order in the $\mu$SR~\cite{Bahrami2021} and NMR results (this work).  Bulk magnetic susceptibility $\chi$ measured using superconducting quantum interference device (SQUID) of sample A reveals a broad local maximum around 14~K, while $\chi$ of sample B increases drastically at low temperatures, as shown in Fig.\ref{knightshift}. The overall magnitude of $\chi$ of sample B is also lower than $\chi$ of sample A at high temperatures.

We also investigated two additional samples with different levels of disorder between those of samples A and B, as indicated by the relative intensity of the split-off NMR peak associated with the unwanted domains with Ag-inclusion. Both of these two extra samples exhibit qualitatively similar behavior as sample B, and hence the details of their NMR properties are beyond the scope of this work.  We note that the change of stoichiometry due to a small amount of Ag inclusion is within the margin of error in energy dispersive X-ray (EDX) analysis (3-5\%), although the relative intensity of the Li NMR side peak reaches $\sim$17\% for the most disordered sample B.  This suggests that each extra Ag site affects multiple Li sites in its vicinity. 

We performed $^7$Li NMR experiments using standard spin echo pulse sequence in a field of 4.5~T, for temperatures ranging from 2~K to 295~K. The $\pi/2$ pulse width ranged from $2-4\:\mu$s and pulse separation time $\tau$ between $\pi/2$ and $\pi$ pulse was $20\:\mu$s. We measured the spin-lattice relaxation rate $1/T_1$ based on inversion recovery, by monitoring the nuclear magnetization $M(t)$ at select delay times $t$. We deduced $1/T_1^{stretch}$ by fitting $M(t)$ against the conventional stretch exponential. Finally, the density distribution $P(1/T_1)$ of $1/T_1$ is determined from $M(t)$ via the ILTT$_1$ analysis technique outlined in the appendix.

\section{NMR Results and Discussion}

\subsection{\label{lslwsec}$^7$Li line shape and width}

\label{lwksec} 

\begin{figure}[t!]
\centering
\includegraphics[width=0.9\linewidth]{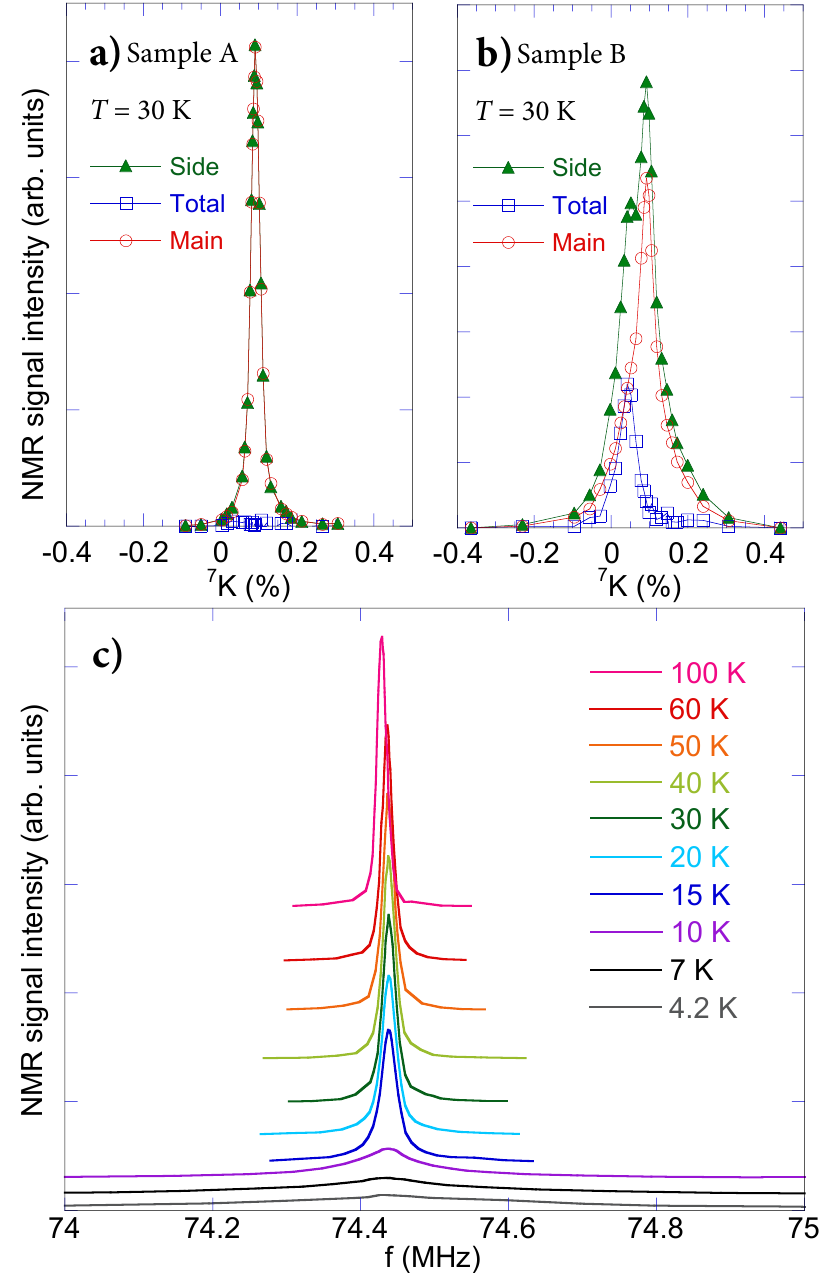}
\caption{NMR line shapes of a) sample A and b) sample B at 30K for the fully relaxed signal (filled triangles), along with the line shapes of the fast (open bullets) and slow (open squares) components. c) cascaded plot of the NMR lineshapes for sample A, observed at temperatures ranging from 4.2~K to 100~K.}
\label{lineshapecascade}
\end{figure}

We compare representative $^7$Li NMR line shapes at 30 K for samples A and B in Fig.\ref{lineshapecascade}(a) and (b). We also show a cascaded plot of all NMR line shapes observed for sample A at 100~K and below in Fig.\ref{lineshapecascade}(c).

There is only one Li site within the honeycomb layer of Ag$_3$LiIr$_2$O$_6$ in Fig.\ref{crystal}, and hence we expect to find only one type of $^7$Li NMR line. In fact, we observed only one sharp $^7$Li NMR peak for sample A, as shown in Fig.\ref{lineshapecascade}(a). However, we can identify a second peak in the total lineshape of sample B, as shown by the green curve of Fig.\ref{lineshapecascade}(b). NMR is a local probe, and hence the presence of the second peak indicates that two different types of $^7$Li sites exist in the disordered sample B.

We can take advantage of the difference in the NMR spin-lattice relaxation rate $1/T_1$ to isolate this second peak. The aforementioned total NMR signal (green triangles in Fig.\ref{lineshapecascade}(a) and (b)) is acquired by waiting for the nuclear spins to return to thermal equilibrium between individual spin echo measurements. This takes between $\sim 200$~ms to $\sim 84$~s depending on the temperature. On the other hand, we can selectively capture the fast component of the signal by repeating the spin echo measurements with 31.3~ms intervals as shown by open red bullets in Fig.\ref{lineshapecascade}(a) and (b). By subtracting the fast component from the total intensity, we obtain the side peak lineshape (blue squares in Fig.\ref{lineshapecascade}(a) and (b)) with very slow relaxation rate.

Sample B exhibits a significant fraction ($\sim 17\%$) of the side peak signal while no side peak is observed in the newest and cleanest sample A.  In view of the Ag inclusion at Li and Ir sites as a cluster, we can attribute the side peak to Li sites within or around the areas of honeycomb layer with these extra Ag sites.

\begin{figure}[t!]
\centering
\includegraphics[width=\linewidth]{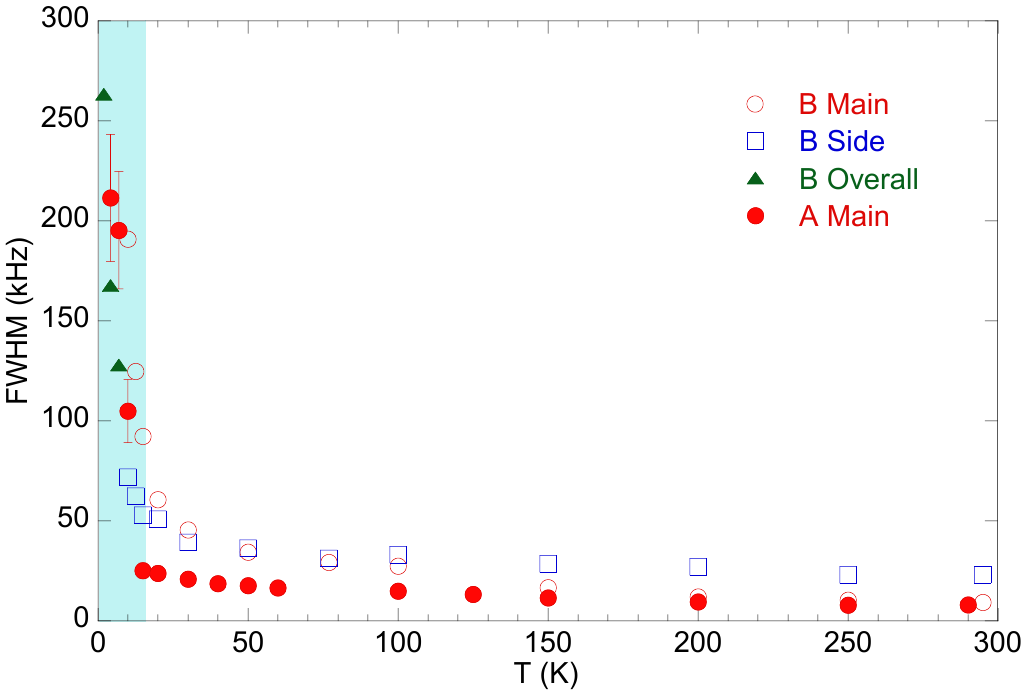}
\caption{FWHM of main peak for both sample A (filled bullets) and sample B (open bullets), along with FWHM of side peak (open squares) and overall width (triangles, below 10~K) of sample B. The FWHM begins to broaden rapidly below $\sim 14$~K and this region is highlighted in light blue.}
\label{FWHM}
\end{figure}

We summarize the full width at half maximum (FWHM) of the lineshapes in Fig.\ref{FWHM}. We see that as temperature decreases below $\sim 14$~K, both the main and side peak broaden rapidly for both samples A and B. The onset of NMR line broadening is generally an indication of the onset of long range ordering or spin freezing. In fact, spin susceptibility data exhibit different behavior between field cooled and zero field cooled measurements starting at 14~K, a typical signature of the onset of spin freezing \cite{Bahrami2021}. This is consistent with our observations. Below 10~K, the main and side peaks of sample B merge and are no longer distinguishable via the fast and slow component of the signal; we show the overall linewidth below 10~K with filled triangles.

\subsection{\label{kssec}Knight shift}

We plot the Knight shift $^7$K of the main and side peak in Fig.\ref{knightshift} along with the spin contribution to bulk magnetic susceptibility $\chi_{spin}$. We deduce $\chi_{spin}$ from the bulk susceptibility $\chi$ using the Van Vleck and diamagnetic contributions such that $\chi = \chi_{spin} + \chi_{dia} + \chi_{vv}$, where $\chi_{dia} \simeq -0.1\times10^{-3}$ emu/mol \cite{Landolt1982} and $\chi_{vv} \simeq 0.14\times10^{-3}$ emu/mol \cite{chaloupkaAPS2013}. Notice that $^7$K of sample A shows nearly identical temperature dependence as $\chi_{spin}$, and exhibits a broad maximum around 25~K. The slightly less suppression of $\chi_{spin}$ observed below 50~K may be attributed to a small amount of paramagnetic defect spin contributions in the bulk susceptibility data.

The inset in Fig.\ref{knightshift} shows $^7$K measured for sample A plotted against $\chi_{spin}$. The main peak Knight shift is linear to magnetic susceptibility $\chi_{spin}$ down to $\sim 100$~K, and a linear fit approximates a chemical shift of $^7$K$_{chem} \simeq 0.018\%$ and the hyperfine coupling constant $A_{hf} = (N_{A}\mu_{B}/N_{n.n.}) \times (d ^7$K$)/(d\chi) = 0.31$~kOe$/\mu_B$ for sample A, where $N_{A}$ is Avogadro's number, $\mu_{B}$ is the Bohr magneton, and $N_{n.n.} = 6$ is the number of nearest neighbor Ir adjacent to each Li site.

The $^7$K of the sample B main peak shows nearly identical behaviour to that of sample A. This indicates that local spin susceptibility of the cleaner parts of sample B far from Ag clusters exhibit nearly identical behavior as clean sample A. On the other hand, $^7$K at the side peak of sample B decreases at lower temperatures and levels off, suggesting that spin susceptibility gets locally suppressed near the Ag clusters. Moreover, the absence of a $^7$K component that increases dramatically below $\sim 10$~K indicates that a greater concentration of unpaired spins are present in sample B and responsible for the steep upturn of bulk $\chi$ at low temperatures.  These findings are consistent with theoretical speculations that disorder in the Kitaev planes may cause appearance of local spin singlets, accompanied by orphaned paramagnetic spins~\cite{Kimchi2018,Knolle_2019}.


\subsection{\label{t1iltsec}Spin dynamics and 1/$T_1$ distribution}

\begin{figure}[h!]
\centering
\includegraphics[width=0.93\linewidth]{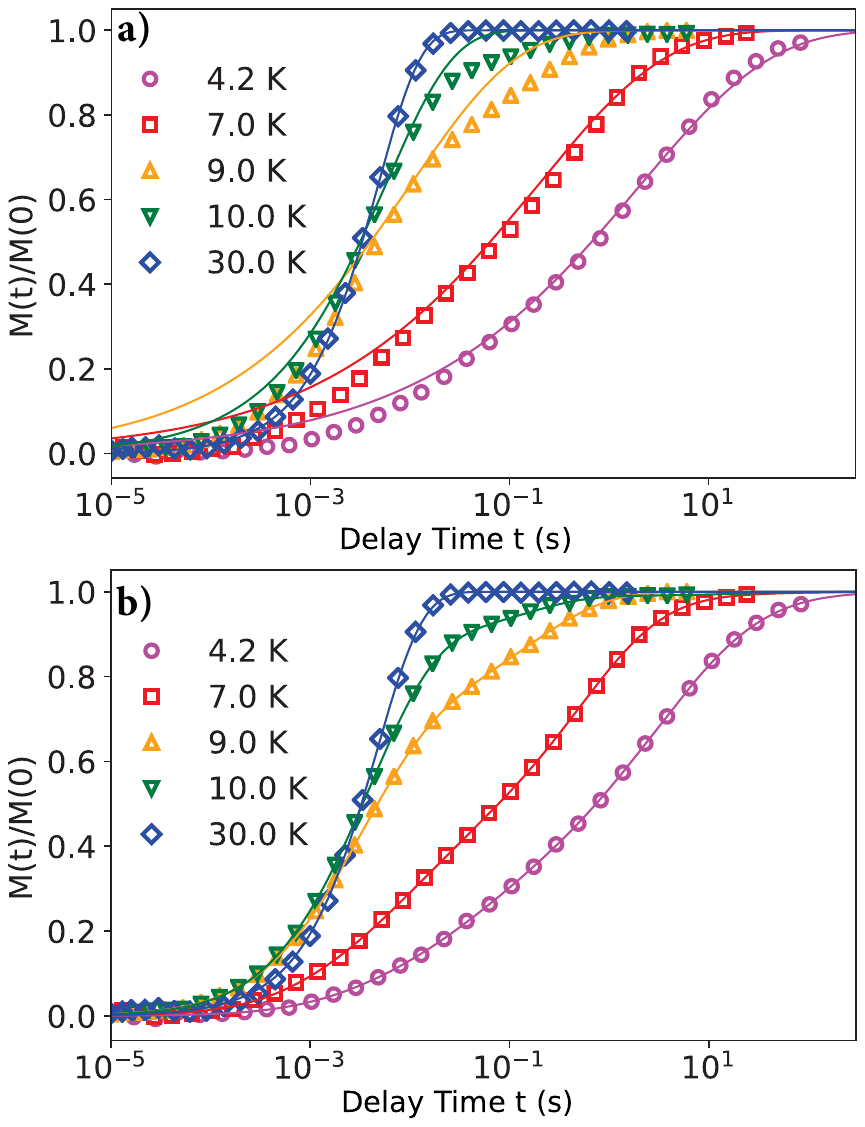}
\caption{The normalized recovery curve $M(t)$ observed for sample A at select temperatures. The solid lines in panel a) show the respective stretch fit results, and the solid lines in panel b) show the ILT fit results. The stretch fit fails to capture all the components when $1/T_1$ becomes highly distributed while the ILT fits the data nicely at all temperatures.}
\label{recovery cascade}
\end{figure}

In order to investigate the Ir spin dynamics, we measured spin-lattice relaxation rate $1/T_1$ at the main peak of sample A and B, and the side peak of sample B. Representative $1/T_1$ recovery curves $M(t)$ at the sample A main peak are plotted in Fig.\ref{recovery cascade}.

To obtain an estimate of $1/T_1$, $M(t)$ is fitted against phenomenological stretched exponential function
\begin{equation}
M(t) = M_0 - A\times\text{exp}(-(t/T_1^{stretch})^\beta),
\label{eq:one}
\end{equation}
where $M_0$, $A$, $T_1^{stretch}$, and $\beta$ are the free parameters. The stretch fit exponent $\beta$ accounts for the distribution of $1/T_1$ with $\beta = 1$ corresponding to no distribution.

\begin{figure}[t!]
\centering
\includegraphics[width=\linewidth]{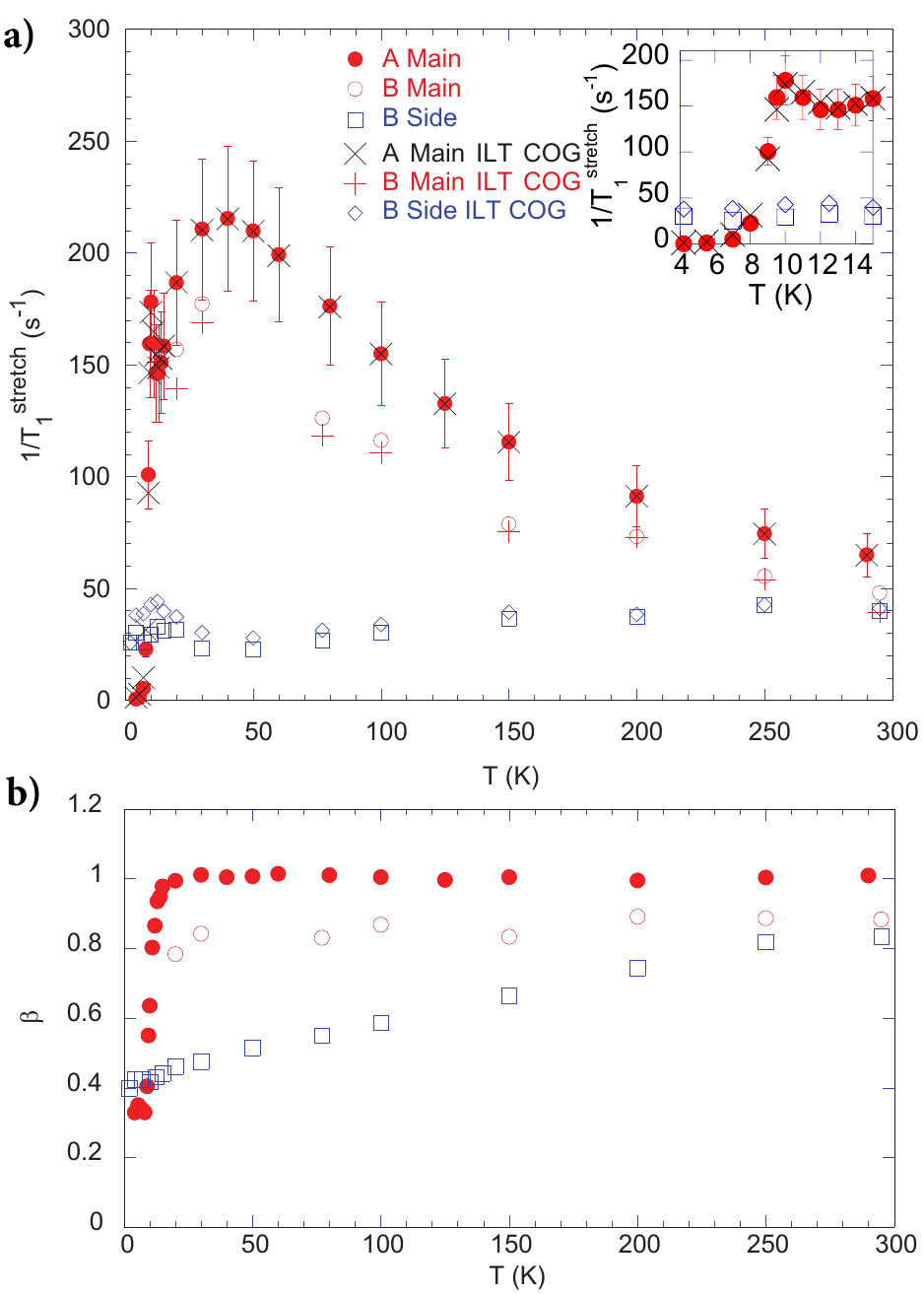}
\caption{a) Spin lattice relaxation rate $1/T_1^{stretch}$ estimated from stretch fit at the main peak of sample A (filled red bullets) and the main (open red circles) and side (open blue squares) peak of sample B. ILT COG of sample A main peak (black X's) and sample B main (red +'s) and side (open blue diamonds) peaks seen in Fig.\ref{cascades} are also plotted. Note that below 10K in sample B, the rapidly broadening main peak is overshadowed by the side peak, and thus $1/T_1$ is measured at the peak of the overall NMR line shape. The inset shows the low temperature data zoomed in to better display the sharp $1/T_1$ peak at $9\pm 1$~K of the sample A main peak. b) The stretch fit parameter $\beta$ for samples A and B. For sample A main peak, we found $\beta = 1$ within experimental uncertainties down to 20~K, implying that there is no distribution in $1/T_1$.}
\label{stretch}
\end{figure}

We summarize the fit results of $1/T_1$ and $\beta$ in Fig.\ref{stretch} along with the center of gravity (COG) of the distribution of $1/T_1$ estimated from  ILT, as discussed later in this section. We see that $1/T_1^{stretch}$ measured at the main peak of sample A and B increases with decreasing temperature and reaches a broad maximum around 40~K. For Sample A, $1/T_1^{stretch}$ then exhibits a second, sharp peak at $T_N = 9\pm1$~K and becomes vanishingly small below it. Similar sharp peaks of $1/T_1$ are generally observed in materials undergoing magnetic long range order, where $1/T_1$ diverges toward the transition temperature due to critical slowing down of spin fluctuations. Our finding is consistent with recent $\mu$SR measurements, where static hyperfine field arising from incommensurate AF order emerges below $\sim 8$~K \cite{Bahrami2021}.  As discussed above in section \ref{lslwsec}, these signatures of long range order are preceded by the aforementioned typical signatures of spin freezing in the bulk $\chi_{spin}$ and NMR line broadening below 14~K, presumably because the residual disorder effects caused by stacking faults are suppressing $T_{N}$ even in cleaner sample A.

The broad peak in the $1/T_1$ of sample A around 40~K is accompanied by the aforementioned broad peak in $^7$K. This finding is similar to the case in Cu$_2$IrO$_3$, where $1/T_1$ and Knight shift at $^{63}$Cu sites also exhibit broad peaks around 40~K but $1/T_1$ does not diverge~\cite{takahashi2019}. Since the suppression of $\chi$ usually signals the short range order of spins in low dimensional systems, it may be perplexing to find that $1/T_1$ is not starting to diverge below $\sim 40$~K. An interesting scenario is that the spin excitation spectrum develops a gap below $\sim 40$~K at a fraction of the Ir-Ir Kitaev interaction $| J_{K} |\sim10^{2}$~K~\cite{Chaloupka_2010}, as expected for Majorana fermions and fluxes \cite{KnolleAPS2015,Yoshitake2016,YoshitakeJul2017,YoshitakeAug2017}. However, the Li atoms are situated at the high symmetry position of the center of six Ir sites. Therefore fluctuating hyperfine magnetic fields arising from incommensurate antiferromagnetic spin fluctuations may nearly cancel out at $^7$Li sites, suppressing $1/T_1$ below 40~K even without a spin excitation gap. An analogous situation was encountered in $^{89}$Y NMR $1/T_1$ data in YBa$_2$Cu$_3$O$_6$ \cite{ohno1990}, and we cannot entirely rule out such a purely geometrical scenario for the observed suppression of $1/T_1$ below $\sim40$~K.

The $1/T_1^{stretch}$ and $\beta$ measured at the main peak of sample B shows roughly the same behavior as sample A with both being slightly smaller. However, $1/T_1^{stretch}$ measured at the side peak of sample B shows qualitatively different behavior as it is much lower and does not peak near 40~K. Instead, $1/T_1^{stretch}$ of the side peak gradually decreases. This is consistent with the gradual decrease of $^7$K of the side peak, and seem consistent with nearly non-magnetic regions that emerge near domains with Ag inclusion.  The emergence of these non-magnetic regions also explain why the magnitude of the the overall bulk susceptibility data in Fig.\ref{knightshift}  is somewhat suppressed for sample B, except in the low temperature region dominated by isolated spins.

$1/T_1^{stretch}$ is generally only a crude approximation of the center of gravity (COG) of the distributed values of $1/T_1$, as we recently demonstrated for various materials \cite{takahashi2019,Singer2020,Arsenault2020}. Accordingly, a more generalized analysis technique is needed to understand the behavior of a highly distributed $1/T_1$ with multiple components.

To get the precise distribution of $1/T_1$, we can apply ILTT$_1$ analysis technique to our $M(t)$ data and deduce the density distribution function of $P(1/T_1)$ \cite{Singer2020,Arsenault2020}. We define $P(1/T_1)$ for a discrete range of relaxation rates $1/T_{1,i}$ as
\begin{equation}
M(t) = \sum_i [1 - 2\times\text{exp}(-t/T_{1,i})]P(1/T_{1,i}) + E(t),
\label{distributiondefinition}
\end{equation}
where $P(1/T_{1,i})$ is the non-negative ILT spectrum weight and $E(t)$ is Gaussian noise.  Incomplete inversion of $M(t)$ is taken into account as explained in the Supplemental Materials of \cite{Singer2020}.   We numerically invert $M(t)$ to obtain $P(1/T_{1})$ based on ILT via the method outlined in appendix \ref{appilt}; see \cite{Singer2020} and its Supplementally Materials for a brief review and the complete details of the ILT procedures.  

\begin{figure}[tb!]
\centering
\includegraphics[width=\linewidth]{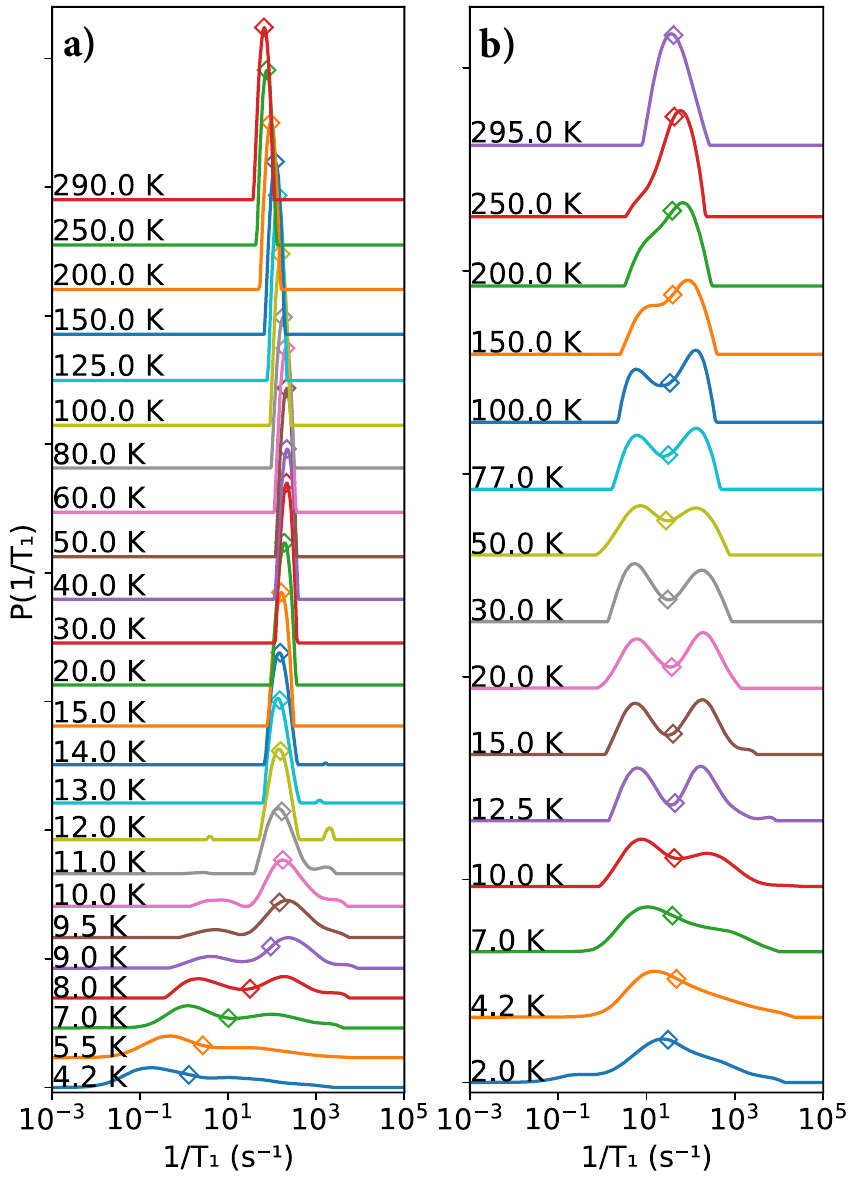}
\caption{Cascaded ILT density distributions $P(1/T_1)$ for a) Sample A main peak and b) Sample B side peak $1/T_1$ relaxation measurements. The distributions are normalized to have equal areas. The temperatures corresponding to each curve are labeled on the left side of the plot. The open diamonds indicate the calculated COG of the ILT distribution. The $P(1/T_1)$ results deduced for $M(t)$ measured at the main peak of sample B (not shown) is similar to the side peak results in panel b), except that the relative intensity of the slower peak of $P(1/T_1)$ is much smaller.}
\label{cascades}
\end{figure}

\begin{figure}[tb!]
\centering
\includegraphics[width=0.95\linewidth]{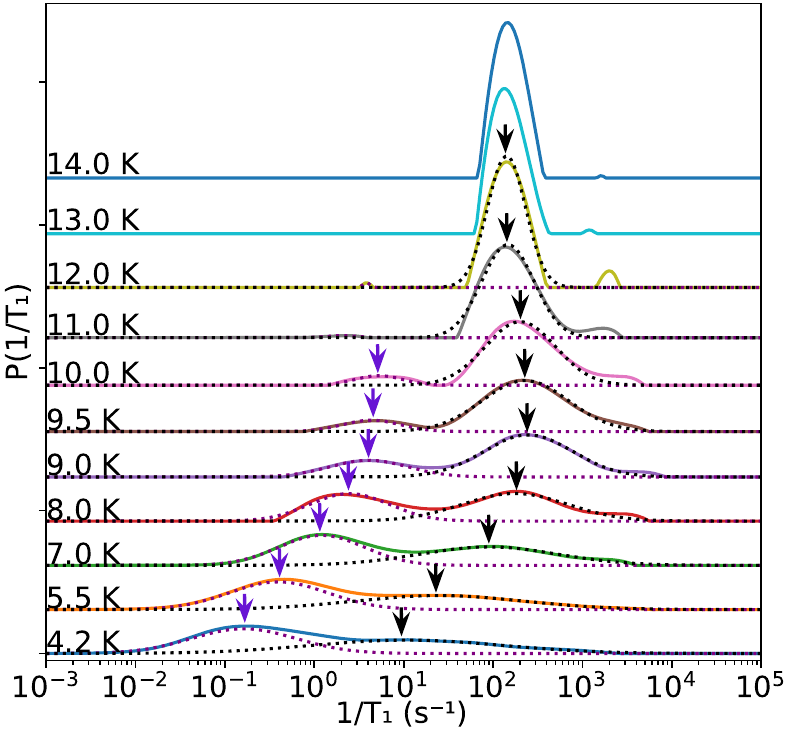}
\caption{Zoomed in view of sample A main peak ILT below 15~K. The black and purple dotted lines below 13~K indicate double Gaussian fits used to deduce the relative fraction of the extra-slow component. The black and purple arrows indicate the peak of the fast and slow components respectively.}
\label{iltzoomedin}
\end{figure}

\begin{figure}[tbh!]
\centering
\includegraphics[width=1.0\linewidth]{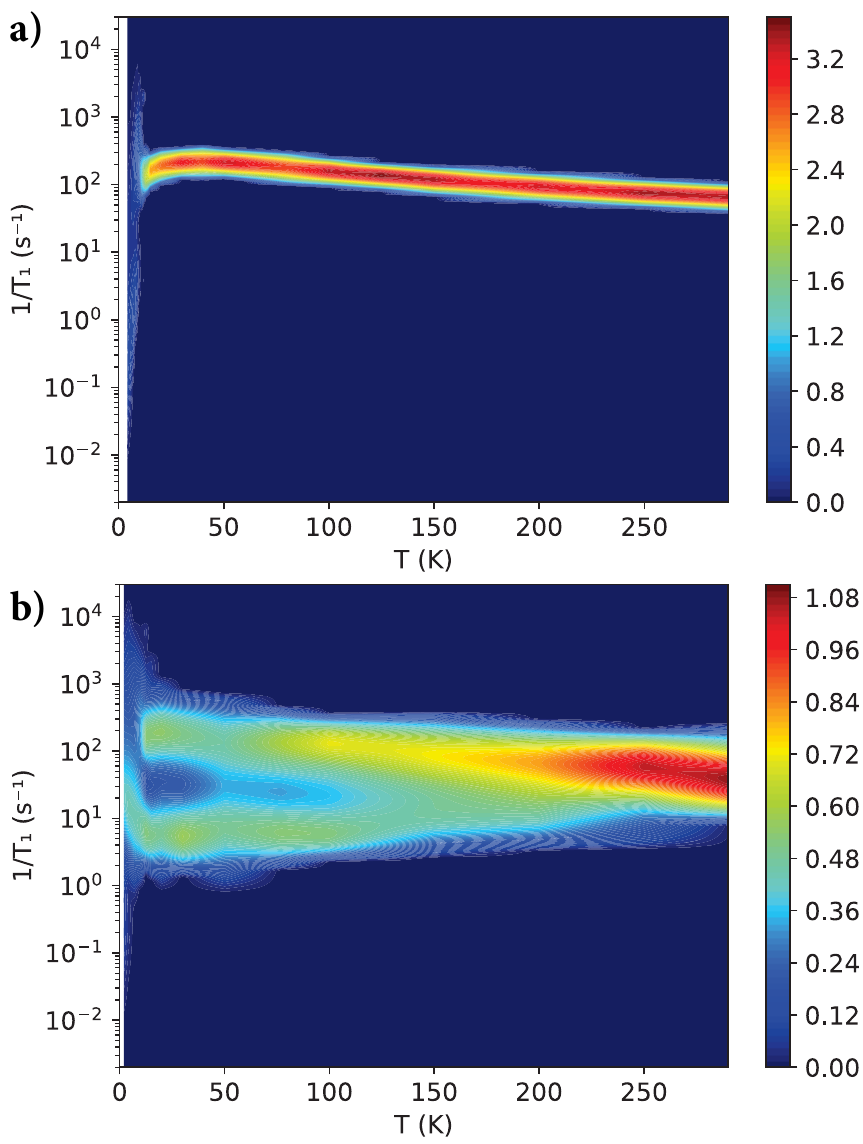}
\caption{Contour plot of the density distribution function $P(1/T_1)$ deduced from $M(t)$ by ILT for a) main peak of sample A and b) side peak of sample B. Note that the faster component in panel b) shows nearly identical behavior as the result in panel a. This means that the fast components observed for the side peak of sample B correspond to the intrinsic behaviour of the material arising from the superposed signals from the main peak.}
\label{contours}
\end{figure}

The ILT approach has major advantages over stretch fit analysis as it is model-independent and thus naturally distinguishes separate components of $1/T_1$ in its distribution.  The total integral of the $P(1/T_1)$ curves is normalized to 1.  Therefore, the area underneath the $P(1/T_1)$ curves between two values of $1/T_1 = a$ and $1/T_1 = b$ represents the fraction of $^7$Li nuclear spins that relaxes with $1/T_1$ values between $a$ and $b$.  We summarize the results for the main peak of sample A in Fig.\ref{cascades}(a) and its color contour plots in Fig.\ref{contours}(a); see Fig.\ref{iltzoomedin} for additional details in the low temperature region.

Looking closely at Fig.\ref{iltzoomedin}, we see that for sample A, some components of $1/T_1$ as marked by black downward arrows indeed peak at $9\pm1$~K with $1/T_1 \approx 200 $~s$^{-1}$, in agreement with the sharp peak observed for $1/T_1^{stretch}$ in Fig.\ref{stretch}(a). In addition, notice that an increasing fraction of $^7$Li sites with $1/T_1$ two orders of magnitude lower emerges at $\sim 10$~K. These Li sites are surrounded by Ir spins whose spin fluctuation time scale has slowed below the NMR frequency ($\sim 74.4$ MHz), owing to static order that is already under way at 10~K. These findings are not revealed by the conventional stretch fit. ILTT$_1$ analysis is better suited to probe these multiple components of $1/T_1$ with qualitatively different behavior.

\begin{figure}[tbh!]
\centering
\includegraphics[width=0.95\linewidth]{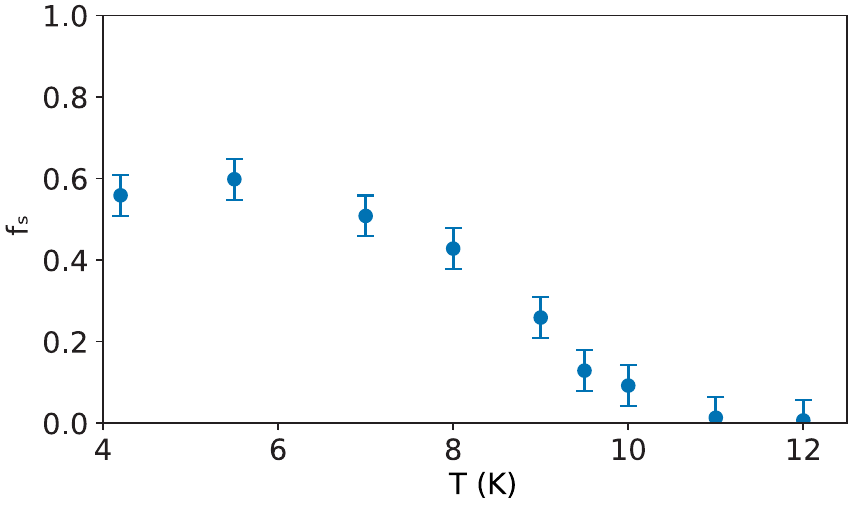}
\caption{Fraction of extra-slow component f$_s$ in $P(1/T_1)$ below 13~K for sample A, estimated by double Gaussian deconvolution of $P(1/T_{1})$ in Fig.\ref{iltzoomedin}.}
\label{staticorder}
\end{figure}

At 4.2~K in sample A, a large fraction of the $P(1/T_1)$ ($\sim 60\%$) appears to be part of the extra-slow component centered around $1/T_{1} \sim$~20 s$^{-1}$ or less, although a substantial fraction ($\sim 40\%$) of $^7$Li nuclear spins still have $1/T_1$ greater than 1 s$^{-1}$. We apply a double Gaussian deconvolution of $P(1/T_1)$ below 13~K (dotted curves in Fig.\ref{iltzoomedin}) to estimate the relative fraction f$_s$ of this slow component, and plot the results in Fig.\ref{staticorder}. From this, we estimate the fraction of static Ir moments f$_s$ at 4.2~K as approximately $60\%$. This explains the significant decrease in $1/T_1^{stretch}$ below 10~K.

For comparison, we present the ILT curves observed at the side peak of sample B in Fig.\ref{cascades}(b), and its color contour plot in Fig.\ref{contours}(b).  Notice that $P(1/T_{1})$ curve splits into two distinct components with $1/T_1$ differing by up to 2 orders of magnitude.  This is simply because the side peak of the NMR lineshape is superposed by the tail of the main peak as shown in Fig.\ref{lineshapecascade}(b), and the latter shows nearly identical behavior as sample A, as readily seen in the similarity between the result in Fig.\ref{contours}(a) and the upper branch of Fig.\ref{contours}(b).  We also confirmed that the ILT results of $P(1/T_{1})$ calculated for the $M(t)$ data measured for sample B at the main peak of the lineshape in Fig.\ref{lineshapecascade}(b) resemble the results in Fig.\ref{cascades}(b), except the relative intensity of the fast component ($1/T_1$ $>$ 10 s$^{-1}$) is much higher; this assures us that the fast component peak seen in Fig.\ref{cascades}(b) indeed arises from the superposed signals of the main peak in the lineshape.  The peak of the slow component in Fig.\ref{cascades}(b) is located at $\sim50$~s$^{-1}$ near room temperature, and quickly drops to the baseline value of $\sim10$~s$^{-1}$ by $\sim100$~K.  Combined with the relative intensity $\sim17$\% in the NMR lineshape in Fig.\ref{lineshapecascade}(b) and the small values of Knight shift in Fig.\ref{knightshift} observed for the side peak, this finding establishes that $\sim17$\% of Li sites are located in a region with suppressed magnetism.  Theoretically, disorder effect in Kitaev lattice are predicted to induce randomly emerging singlets accompanied by isolated free spins, generally referred to as orphaned spins in the literature \cite{Kimchi2018,Knolle_2019}.  If we attribute the side peak to such singlets, we can estimate their excitation gap to be of the oder of $\sim100$~K in this material.  The significant volume of nearly non-magnetic domains accompanied by orphaned spins explains why the clear signatures of magnetic long range order can be easily masked by disorder. 

\section{Conclusions}

We have probed the intrinsic spin susceptibility and spin dynamics of a clean sample A of Ag$_3$LiIr$_2$O$_6$ using $^7$Li NMR. The sharp peak in $1/T_1^{stretch}$ along with the behavior of the very fast component in the ILT is consistent with $\sim 60\%$ of the sample exhibiting long-range ordering starting $T_N = 9\pm1$~K. However, the ILT analysis of the highly distributed $1/T_1$ revealed that some parts are already entering the ordered phase below $\sim 10$~K while some other fraction is still fluctuating even at 4.2~K for nominally defect-free sample A.  This is presumably due to the residual disorder effects arising from stacking faults revealed in TEM measurements.

To elucidate the influence of structural disorder, we also compare the NMR results for highly disordered sample B with clusters of unwanted Ag occupying at Li and Ir sites in addition to stacking faults.  We demonstrated that these domains with Ag inclusion give rise to a new side peak in the NMR lineshape with suppressed Knight shift and $1/T_1$, indicating that disorder locallys suppresses magnetism. 

The highly disordered sample of Ag$_3$LiIr$_2$O$_6$ is similar to other Kitaev materials such as Cu$_2$IrO$_3$ and H$_3$LiIr$_2$O$_6$ in that there is no definitive evidence of long range order, such as diverging $1/T_1$ \cite{takahashi2019,Kitagawa2018}. Absence of such clear-cut signature of long range order led to earlier proposals for the quantum spin liquid ground state in these materials. But our findings for disordered sample B here and elsewhere \cite{Bahrami2021} indicate that disorder could easily mask the signature of long range order. 

\begin{acknowledgments}
The work at McMaster was supported by NSERC. The work at Boston College was supported by the National Science Foundation under DMR-1708929. P.M.S. is supported by The Rice University
Consortium for Processes in Porous Media.

\end{acknowledgments}


\appendix

\section{Inverse Laplace Transform}\label{appilt}

We can numerically deduce the density distribution function $P(1/T_1)$ in eq.(\ref{distributiondefinition}) using ILTT1 analysis technique without relying on phenomenological models by computationally inverting $M(t)$. This was done previously in Cu$_2$IrO$_3$, La$_{1.875}$Ba$_{0.125}$CuO$_6$, and La$_{1.885}$Sr$_{0.115}$CuO$_4$ and had revealed information beyond what was shown by the stretch fit and 2-component fit \cite{takahashi2019,Singer2018,Singer2020,Arsenault2020}.

Here, we provide an outline of ILTT1 analysis. For a discrete range of time steps $\{t_i\}$, we can reduce eq.(\ref{distributiondefinition}) to its vector form
\begin{equation}
\textbf{M} = K\textbf{P} + \textbf{E}, \qquad \textbf{P} = \{P(1/T_{1,i}) \geq 0\},
\label{linearequation}
\end{equation}
where $K$ is the kernel matrix and $\textbf{E}$ is a vector representing Gaussian noise. For a sufficient number of $1/T_{1,i}$, this is an ill-posed problem with non-unique solutions that are sensitive to noise \cite{Mitchell2012,asthgiri2020}. We thus use Tikhonov regularization to introduce a smoothing parameter $\alpha$, such that we can find the unique solution $\textbf{P}$ under the constraint $\textbf{P} \geq 0$ which minimizes the cost function
\begin{equation}
\begin{split}
\textbf{P} = \underset{\textbf{P} \geq 0}{\text{arg min}} \, |\textbf{M} - K\textbf{P}|^2 + \alpha|\textbf{P}|^2,
\end{split}
\label{smooth min}
\end{equation}
where $|...|$ is the vector norm. To prevent over-fitting (the ILT distribution becomes greatly affected by the noise) and under-fitting (important information in $M(t)$ becomes lost), $\alpha$ is chosen such that the ILT fit deviates from $M(t)$ in proportion to the experimental noise. We refer readers to section II of \cite{Singer2020} and its Appendix for more details.

\bibliographystyle{apsrev4-2}


%


\end{document}